\documentclass[12pt]{article}
\usepackage{amsmath,amssymb,indentfirst,bm}
\usepackage{graphicx, epsfig}
\usepackage[authoryear]{natbib}
\usepackage{amsthm, verbatim}
\usepackage{setspace}
\usepackage{hyperref}

\title{A statistical test to identify differences in clustering structures}

\author{Andr\'e Fujita*\\{\small Department of Computer Science, Institute of Mathematics and Statistics,}\\ {\small University of S\~ao Paulo, Brazil.}\\
Daniel Y. Takahashi\\{\small Department of Psychology and Neuroscience Institute, Green Hall,}\\ {\small Princeton University, USA.}\\
Alexandre G. Patriota\\{\small Department of Statistics, Institute of Mathematics and Statistics,}\\ {\small University of S\~ao Paulo, Brazil.}\\
Jo\~ao R. Sato\\{\small Center of Mathematics, Computation, and Cognition, Federal University of ABC, Brazil.}\\\\
{\small {*Corresponding author}}\\
{\small Rua do Matão, 1010 - Building C, Cidade Universit\'aria}\\ 
{\small S\~ao Paulo - SP - Brasil - CEP 05508-090}\\
{\small Phone: +55 11 3091 5177}
}

\date{}

\begin{document}
\maketitle

%\doublespacing
\begin{abstract}
Statistical inference on functional magnetic resonance imaging (fMRI) data is an important task in brain imaging. One major hypothesis is that the presence or not of a psychiatric disorder can be explained by the differential clustering of neurons in the brain. In view of this fact, it is clearly of interest to address the question of whether the properties of the clusters have changed between groups of patients and controls. The normal method of approaching group differences in brain imaging is to carry out a voxel-wise univariate analysis for a difference between the mean group responses using an appropriate test (e.g. a t-test)  and to assemble the resulting ``significantly different voxels'' into clusters, testing again at cluster level. In this approach of course, the primary voxel-level test is blind to any cluster structure. Direct assessments of differences between groups (or reproducibility within groups) at the cluster level have been rare in brain imaging. For this reason, we introduce a novel statistical test called ANOCVA - ANalysis Of Cluster structure Variability, which statistically tests whether two or more populations are equally clustered using specific features. The proposed method allows us to compare the clustering structure of multiple groups simultaneously, and also to identify features that contribute to the differential clustering. We illustrate the performance of ANOCVA through simulations and an application to an fMRI data set composed of children with ADHD and controls. Results show that there are several differences in the brain's clustering structure between them, corroborating the hypothesis in the literature. Furthermore, we identified some brain regions previously not described, generating new hypothesis to be tested empirically.

\vspace{0.1cm}
\noindent{\it Keywords}: clustering; silhouette method; statistical test.
\end{abstract}
%----------------------------------------- SECAO 1 --------------------------------------

\section{Introduction}\label{introduction}

Biological data sets are growing enormously, leading to an information-driven science \citep{Stein08} and allowing previously impossible breakthroughs. However, there is now an increasing constraint in identifying relevant characteristics among these large data sets. For example, in medicine, the identification of features that characterize control and disease subjects is key for the development of diagnostic procedures, prognosis and therapy \citep{Rubinov10}. Among several exploratory methods, the study of clustering structures is a very appealing candidate method, mainly because several biological questions can be formalized in the form: Are the features of populations A and B equally clustered? One typical example occurs in neuroscience. It is believed that the brain is organized in clusters of neurons with different major functionalities, and deviations from the typical clustering pattern can lead to a pathological condition \citep{Grossberg00}. Another example is in molecular biology, where the gene expression clustering structures depend on the analyzed population (control or tumor, for instance) \citep{Furlan11, Wang13}. Therefore, in order to better understand diseases, it is necessary to differentiate the clustering structures among different populations. This leads to the problem of how to statistically test the equality of clustering structures of two or more populations followed by the identification of features that are clustered in a different manner. The traditional approach is to compare some descriptive statistics of the clustering structure (number of clusters, common elements in the clusters, etc) \citep{Meila07, Cecchi09, Kluger03}, but to the best of our knowledge, little or nothing is known regarding formal statistical methods to test the equality of clustering structures among populations. With this motivation, we introduce a new statistical test called ANOCVA - ANalysis Of Cluster structure VAriability - in order to statistically compare the clustering structures of two or more populations. 

Our method is an extension of two well established ideas: the silhouette statistic \citep{Rousseeuw87}  and  ANOVA. Essentially, we use the silhouette statistic to measure the ``variability" of the clustering structure in each population. Next, we compare the silhouette among populations. The intuitive idea behind this approach is that we assume that populations with the same clustering structures also have the same ``variability''. This simple idea allows us to obtain a powerful statistic test for equality of clustering structures, which  (1) can be applied to a large variety of clustering algorithms; (2) allows us to compare the clustering structure of multiple groups simultaneously;  (3) is fast and easy to implement; and (4) identifies features that significantly contribute to the differential clustering.

We illustrate the performance of ANOCVA through simulation studies under different realistic scenarios and demonstrate the power of the test in identifying small differences in clustering among populations. We also applied our method to study the whole brain functional magnetic resonance imaging (fMRI) recordings of 759 children with typical development (TD), Attention deficit hyperactivity disorder (ADHD) with hyperactivity/impulsivity and inattentiveness, and ADHD with hyperactivity/impulsivity without inattentiveness. ADHD is a psychiatric disorder that usually begins in childhood and often persists into adulthood, affecting at least 5-10\% of children in the US and non-US populations \citep{Fair07}. Given its prevalence, impacts on the children's social life, and the difficult diagnosis, a better understanding of its pathology is fundamental. The statistical analysis using ANOCVA on this large fMRI data set composed of ADHD and subjects with TD identified brain regions that are consistent with already known literature of this physiopathology. Moreover, we have also identified some brain regions previously not described as associated with this disorder, generating new hypothesis to be tested empirically.

%The article is organized as follows. The required definitions and statistical tests are described in Section \ref{methods}. Simulations under several scenarios are constructed to evaluate the statistical power of the method in Section \ref{simulation}. The proposed approach is applied to a large fMRI data set in order to identify ROIs that are associated with ADHD in Section \ref{adhd}. A discussion is provided in Section \ref{final}.

\section{Methods}
\label{methods}

We can describe our problem in the following way. Given $k$ populations $T_{1}, T_{2}, \ldots, T_{k}$ where each population $T_{j}$ ($j=1, \ldots, k$), is composed of $n_{j}$ subjects, and each subject has $N$ items that are clustered in some manner, we would like to verify whether the cluster structures of the $k$ populations are equal and, if not, which items are differently clustered. To further formalize our method, we must define what we mean by cluster structure. The silhouette statistic is used in our proposal to identify the cluster structure. We briefly describe it in the next section.

\subsection{The silhouette statistic}\label{silhouette}
The silhouette method was proposed in 1987 by \cite{Rousseeuw87} with the purpose of verifying whether a specific item was assigned to an appropriate cluster. In other words, the silhouette statistic is a measure of goodness-of-fit of the clustering procedure. Let $\mathcal{X} = \{x_1, \ldots, x_N\}$ be the items of one subject that are clustered into $\mathcal{C} = \{C_1, \ldots, C_r\}$ clusters by a clustering algorithm according to an optimal criterion. Note that $\mathcal{X} = \bigcup_{q=1}^r C_q$. Denote by $d(x, y)$ the dissimilarity (e.g. Euclidian, Manhattan, etc) between items $x$ and $y$ and define 
\begin{equation}\label{Euclidian-metric}
	d(x, C) =  \frac{1}{\#C} \sum_{y \in C} d(x,y)
\end{equation}
as the average dissimilarity of $x$ to all items of cluster $C \subset \mathcal{X}$ (or $C \in \mathcal{C}$), where  $\# C$ is the number of items of $C$. Denote by $D_q \in \mathcal{C}$ the cluster to which $x_q$ has been assigned by the clustering algorithm and by $E_q \in \mathcal{C}$ any other cluster different of $D_q$, for all $q=1, \ldots, N$. All quantities involved in the silhouette statistic are given by  
\[a_{q} = d(x_{q},D_q)  \quad \mbox{and} \quad b_{q} = \min_{E_q \neq D_q} d(x_{q}, E_q), \quad \mbox{for} \ q=1,\ldots, N,
\]
where $a_{q}$ is the ``within'' dissimilarity and $b_{q}$ is the smallest ``between'' dissimilarity for the sample unit $x_q$. Then a natural proposal to measure how well item $x_{q}$ has been clustered is given by the silhouette statistic\citep{Rousseeuw87}
\begin{equation}\label{s-stat}
	s_{q} = \left\{
\begin{array}{ll}
	\dfrac{b_{q} - a_{q}}{\max\{b_{q}, a_{q}\}}, & \mbox{ if } \# D_q > 1,\\
0, & \mbox{ if } \# D_q = 1.
\end{array}\right.
\end{equation}

The choice of the silhouette statistic is interesting due to its interpretations. Notice that, if $s_{q} \approx 1$, this implies that the ``within'' dissimilarity is much smaller than the smallest ``between'' dissimilarity ($a_{q} \ll b_{q}$). In other words, item $x_q$ has been assigned to an appropriate cluster since the second-best choice cluster is not nearly as close as the actual cluster. If $s_{q} \approx 0$, then $a_{q}\approx b_{q}$, hence it is not clear whether $x_q$ should have been assigned to the actual cluster or to the second-best choice cluster because it lies equally far away from both. If $s_{q} \approx -1$, then $a_{q} \gg b_{q}$, so item $x_q$ lies much closer to the second-best choice cluster than to the actual cluster. Therefore it is more natural to assign item $x_q$ to the second-best choice cluster instead of the actual cluster because this item $x_q$ has been ``misclassified''. To conclude, $s_{q}$ measures how well item $x_q$ has been labeled. 

%It is important to note that we can attain the quantities $a_q$ and $b_q$ through a matrix of distances and a vector of labels

Let ${\bf Q} = \{d(x_l, x_q)\}$ be the $(N \times N)$-matrix of dissimilarities, then it is symmetric and has zero diagonal elements. Let ${\bf l} = (l_1, l_2, \ldots, l_N)$ be the labels obtained by a clustering algorithm applied to the dissimilarity matrix ${\bf Q}$, i.e., the labels represent the cluster each item belongs to. It can be easily verified that the dissimilarity matrix ${\bf Q}$ and the vector of labels ${\bf l}$ are sufficient to compute the quantities $s_1, \ldots, s_N$. In order to avoid notational confusions, we will write $s_q^{({\bf Q}, {\bf l})}$ rather than $s_q$ for all $q=1, \ldots, N$, because we deal with many data sets in the next section.

\subsection{Extension of the silhouette approach}

In the previous section, we introduced notations when we have $N$ items in one subject. In the present section, we extend the approach to many populations and many subjects in each population. Let $T_1, T_2, \ldots, T_k$ be $k$ types of populations. For the $j$th population, $n_j$ subjects are collected, for $j=1,\ldots, k$. In order to establish notations, the items of the $i$th subject taken from the $j$th population are represented by the matrix ${\bf X}_{i,j} = ({\bf x}_{i,j,1}, \ldots, {\bf x}_{i,j,N})$ where each item ${\bf x}_{i,j,l}$ $(l=1,\ldots, N)$ is a vector.% In a more general application, each ${\bf x}_{i,j,l}$ can also be a vector containing the spatial position in the cartesian space of an item that will be clustered, for instance.

%It should be mentioned that the vectors $X_{lj}$ and $X_{qj}$ may contain measures over time, therefore their elements are not necessarily synchronized and direct operations between these vector are not allowed (averages, matrix of covariances and so on).

First we define the matrix of dissimilarities among items of each matrix ${\bf X}_{i,j}$, by 
\[{\bf A}_{i,j} = d({\bf x}_{i,j,l}, {\bf x}_{i,j,q}), \quad \mbox{for } i=1,\ldots, n_j, \quad j=1,\ldots, k.\]

Notice that each ${\bf A}_{i,j}$ is symmetric with diagonal elements equal zero. Also, we define the following average matrices of dissimilarities 
\[
\bar{{\bf A}}_{j} = \frac{1}{n_j} \sum_{i=1}^{n_j} {\bf A}_{i,j} =  \frac{1}{n_j}\sum_{i=1}^{n_j} d({\bf x}_{i,j,l}, {\bf x}_{i,j,q}) \quad \mbox{and}\quad
\bar{\bar{{\bf A}}} = \frac{1}{n}\sum_{j=1}^k n_j \bar{{\bf A}}_j
\]
where $n = \sum_{j=1}^k n_j$, $l,q=1,\ldots, N$. The ($N \times N$)-matrices $\bar {\bf A}_1, \ldots, \bar {\bf A}_k$ and $\bar{\bar{{\bf A}}}$ are the only quantities required for proceeding with our proposal.

%Assume that $E(d(x_{ijl}, x_{ijq})^2)< \infty$, then each $\bar{A}_j$ converges almost sure to 

Now, based on the matrix of dissimilarities $\bar{\bar{\bf{A}}}$ we can use a clustering algorithm to find the clustering labels ${\bf l}_{\bar{\bar{{\bf A}}}}$. Then, we compute the following silhouette statistics 
\[s_q^{(\bar{\bar{{\bf A}}},{\bf l}_{\bar{\bar{{\bf A}}}})} \quad \mbox{and} \quad s_q^{(\bar{{\bf A}}_j,{\bf l}_{\bar{\bar{{\bf A}}}})}, \quad \mbox{for} \ q=1, \ldots, N.\]

The former is the silhouette statistic based on the matrix of dissimilarities $\bar{\bar{{\bf A}}}$ and the latter is the silhouette statistic based on the dissimilarity matrix $\bar{{\bf A}}_j$, both obtained by using the clustering labels computed via the matrix $\bar{\bar{{\bf A}}}$. We expect that if the items from all populations $T_1, \ldots, T_k$ are equally clustered, the quantities $s_q^{(\bar{\bar{{\bf A}}},{\bf l}_{\bar{\bar{{\bf A}}}})}$ and $s_q^{(\bar{{\bf A}}_j,{\bf l}_{\bar{\bar{{\bf A}}}})}$ must be close for all $j=1, \ldots k$ and $q=1, \ldots, N$.

\subsection{Statistical tests}\label{test}

Define the following vectors

\[{\bf S} =\left(s_1^{(\bar{\bar{{\bf A}}},{\bf l}_{\bar{\bar{\bf {A}}}})} , \ldots, s_N^{(\bar{\bar{{\bf A}}},l_{\bar{\bar{{\bf A}}}})} \right)^{\top} \quad \mbox{and} \quad {\bf S}_j =\left(s_1^{(\bar{{\bf A}}_j,{\bf l}_{\bar{\bar{{\bf A}}}})} , \ldots, s_N^{(\bar{{\bf A}}_j,{\bf l}_{\bar{\bar{{\bf A}}}})}\right)^{\top}.
\]

We want to test if all $k$ populations are clustered in the same manner, i.e.:

\begin{quote}
$\text{H}_0:$ ``Given the clustering algorithm, the data from $T_{1}, T_{2}, \ldots, T_{k}$ are equally clustered''.\\
{\it versus}\\
$\text{H}_1:$``At least one is clustered in a different manner''. 
\end{quote} 

where the test statistic is defined by
\[
	\Delta S = \sum_{j=1}^k ({\bf S} - {\bf S}_j)^\top ({\bf S} - {\bf S}_j).
\]

In other words, given the clustering structure of the items of each subject, we would like to test if the items are equally clustered among populations. 

Now, suppose that the null hypothesis is rejected by the previous test. Thus, a natural next step is to identify which item is clustered in a different manner among populations. This question can be answered by analyzing each $s_{q}$ with the following statistical test

\begin{quote}
$\text{H}_0:$ ``Given the clustering algorithm, the $q$th item ($q=1, \ldots, N$) is equally clustered among populations''.\\
{\it versus}\\
$\text{H}_1:$ ``The $q$th item is not equally clustered among populations''.
\end{quote} 

where the test statistic is defined by $\Delta s_{q}=\left(s_{q}^{(\bar{\bar{{\bf A}}},{\bf l}_{\bar{\bar{{\bf A}}}})} - \frac{1}{k} \sum_{j=1}^{k}s_{q}^{(\bar{{\bf A}}_j,{\bf l}_{\bar{\bar{{\bf A}}}})}\right)^2$, for $q=1,\ldots, N$.

The exact or asymptotic distributions of both $\Delta S$ and $\Delta s_{q}$ are not trivial, therefore, we use a computational procedure based on bootstrap \citep{Efron94} to construct the empirical null distributions.

The bootstrap implementation of both tests is as follows:

\begin{enumerate}
\item Resample with replacement $n_j$ subjects from the entire data set $\{T_{1}, T_{2}, \ldots, T_{k}\}$ in order to construct bootstrap samples $T_{j}^{*}$, for $j=1,\ldots, k$. 

\item Calculate $\bar{{\bf A}}_{j}^{*}$, $\bar{\bar{{\bf A}}}^{*}$, $s_{q}^{(\bar{\bar{{\bf A}}},{\bf l}_{\bar{\bar{{\bf A}}}})*}$ and $s_{q}^{(\bar{{\bf A}},{\bf l}_{\bar{\bar{{\bf A}}}})*}$, for $q=1,\ldots, N$, using the bootstrap samples $T_{j}^{*}$.

\item Calculate $\hat{\Delta S}^{*}$ and $\hat{\Delta s}_{q}^{*}$.

\item Repeat steps 1 to 3 until obtaining the desired number of bootstrap replications.

\item The p-values from the bootstrap tests based on the observed statistics $\Delta S$ and $\Delta s_{q}$ are the fraction of replicates of $\hat{\Delta S}^{*}$ and $\hat{\Delta s_{q}}^{*}$ on the bootstrap data set $T_{j}^{*}$, respectively, that are at least as large as the observed statistics on the original data set.

\end{enumerate}

The data analysis can be described as shown in 

\begin{figure}[!t]
\centering
\includegraphics[angle=0, width=5.5in]{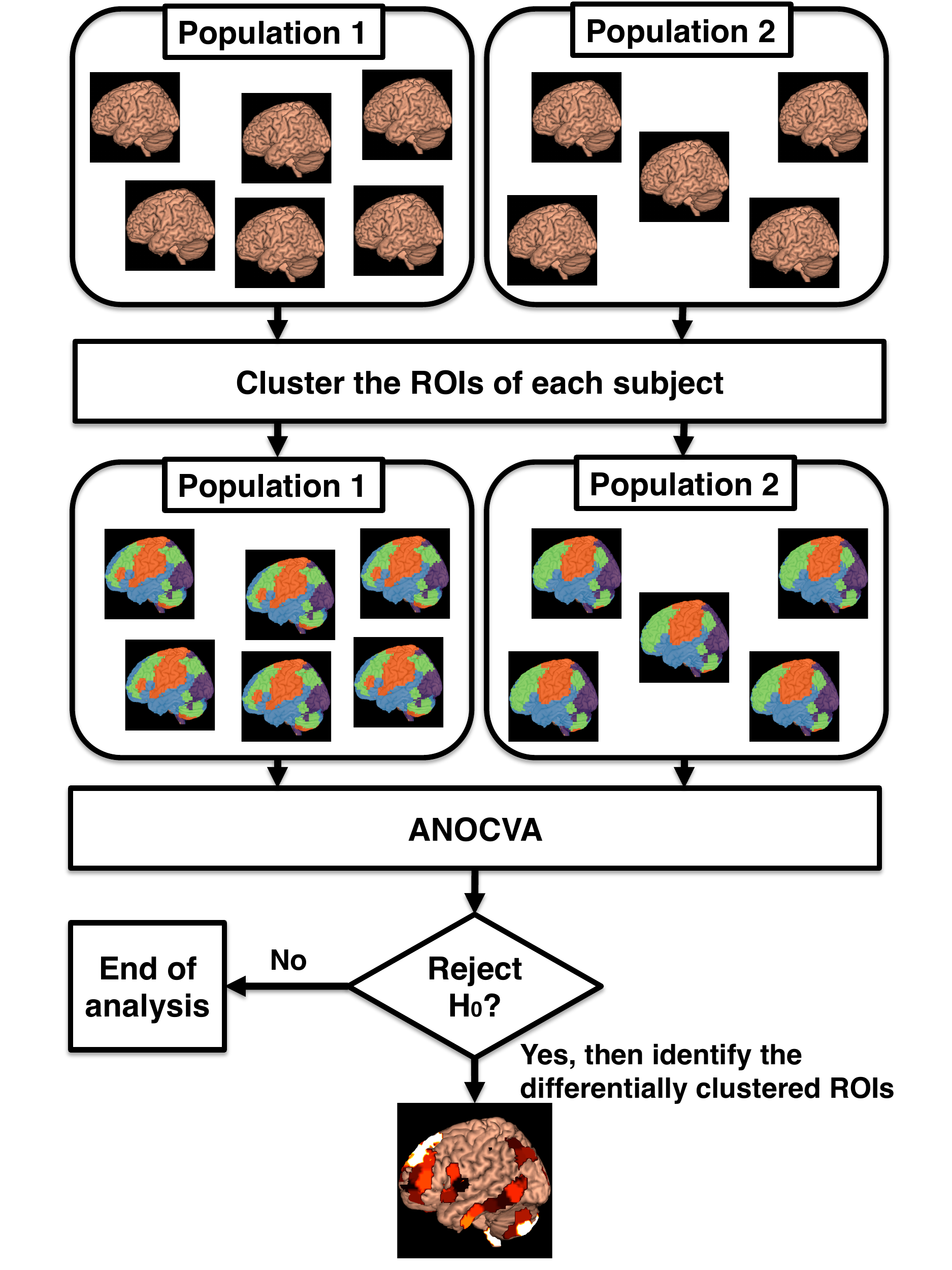}
\caption{{\bf Data analysis pipeline.} The analysis consists in clustering the regions-of-interest (ROIs) of each subject and then, testing by ANOCVA (analyzing the test statistic $\Delta S$) whether the ROIs are equally clustered between populations. If they are not equally clustered, i.e., the null hypothesis ($\text{H}_{0}$) is rejected, the ROIs that most contribute to this differential clustering can be identified by using the test statistic $\Delta s_{q}$.}
\label{fig:test-schema}
\end{figure}

 \ref{fig:test-schema}. 

%%%%%%%%%%%%%%%%%%%%
%% Figure
%%%%%%%%%%%%%%%%%%%%

\subsection{Simulation description}\label{simulation}
\begin{figure}[!t]
\centering
\includegraphics[angle=0, width=5.5in]{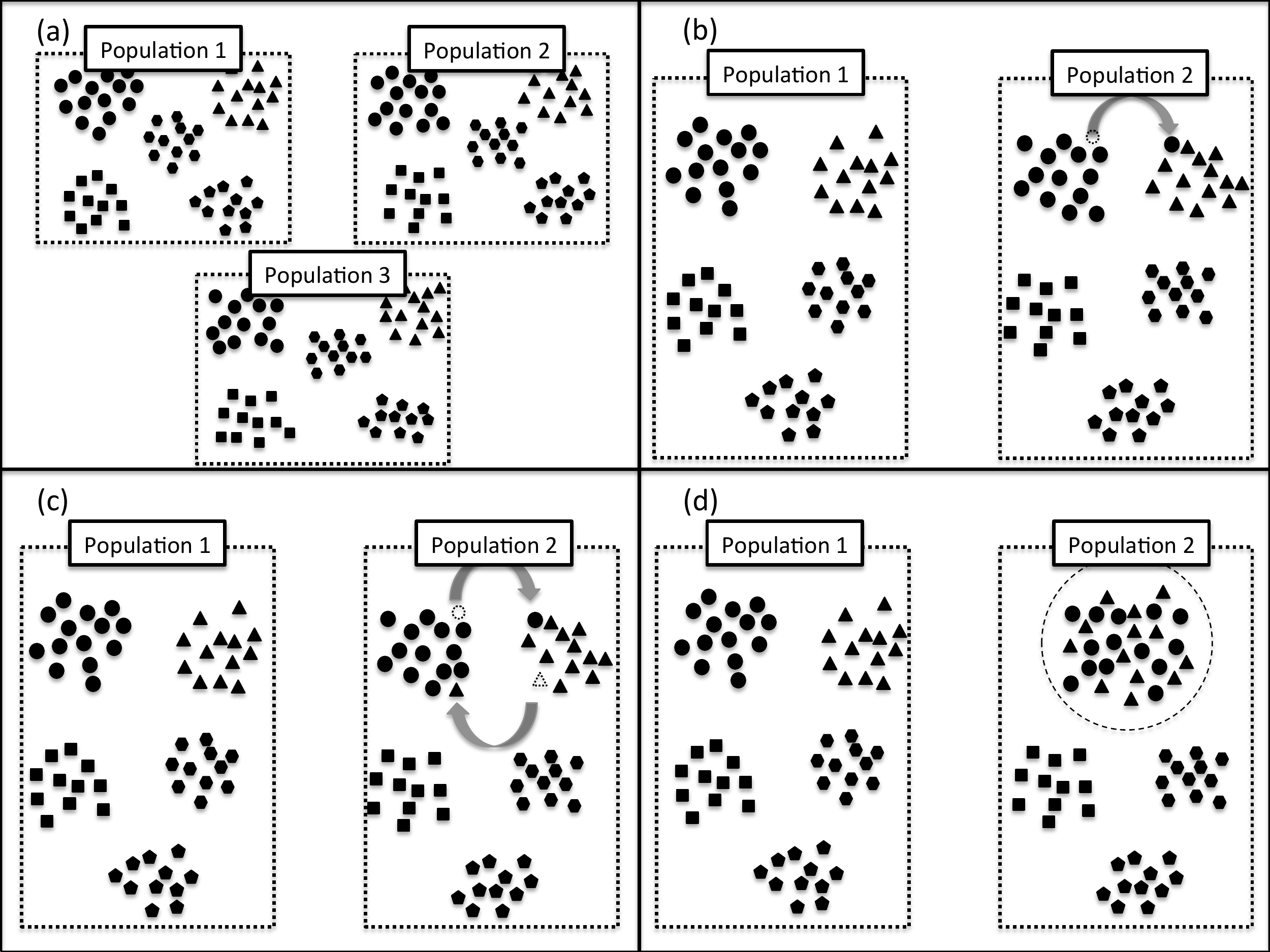}
\caption{{\bf Caricatural illustration of the simulations.} (a) clusters under the null hypothesis, i.e., the clustering structure is the same in all populations (scenario (i)); (b) one item moves to another cluster, i.e., the number of items of two clusters change (scenario (ii)); (c) a swap of items between two clusters, i.e., the clustering structure changes but not the number of items of each cluster (scenario (iii)); (d) merge of clusters, i.e., the number of clusters change between populations (scenario (iv)).}
\label{fig:simulacao}
\end{figure}

Four scenarios were designed to validate some features of ANOCVA, such as size and power of the proposed tests. Our first scenario evaluates the size of the test while the second, third, and forth ones evaluate the power under different settings: (i) all items are equally clustered along three populations ($k=3$) -- this is our null hypothesis (Figure \ref{fig:simulacao}a); (ii) when one single item of cluster A in population 1 is labeled as cluster B in population 2 ($k=2$) -- in this alternative hypothesis, the number of items inside the clusters changes (Figure \ref{fig:simulacao}b); (iii) when one single item of cluster A in population 1 is labeled as cluster B in population 2, and another item from cluster B in population 1 is labeled as cluster A in population 2 ($k=2$) -- in this alternative hypothesis, the number of items inside the clusters does not change (Figure \ref{fig:simulacao}c); and (iv) when two clusters in population 1 are grouped into one single cluster in population 2 ($k=2$) -- in this alternative hypothesis, the number of clusters between populations changes (Figure \ref{fig:simulacao}d).

Each population is composed of 20 subjects ($n_{j}=20$, for $j=1,2,3$). Each subject is composed of $N=100$ items (${\bf X}_{i,j}=({\bf x}_{i,j,1},\ldots,{\bf x}_{i,j,100})$). The items are generated by normal distributions with unit variance. We assume that items generated by the same distribution belong to the same cluster.

The construction of the four scenarios were carried out in the following manner:
\begin{enumerate}

\item Scenario (i): items (${\bf x}_{i,j,1},\ldots, {\bf x}_{i,j,20}$), (${\bf x}_{i,j,21},\ldots, {\bf x}_{i,j,40}$), (${\bf x}_{i,j,41},\ldots, {\bf x}_{i,j,60}$), (${\bf x}_{i,j,61},\\\ldots, {\bf x}_{i,j,80}$), and (${\bf x}_{i,j,81},\ldots, {\bf x}_{i,j,100}$) for $i=1, \ldots, 20$ and $j=1,2,3$ are centered at positions (0, 0), (2, 2), (4, 4), (6, 6), and (8, 8), respectively. This scenario represents the configuration of three populations, each population composed of 20 subjects, where each subject is composed of 100 items that are clustered in five groups. Notice that the items of the subjects of the three populations are equally clustered, i.e., they are under the null hypothesis.

\item Scenario (ii): items of the $i$th ($i=1,\ldots,20$) subject taken from the $j$th ($j=1,2$) population are generated in the same manner as in scenario (i), except by item ${\bf x}_{i,2,1}$, which is centered at position (2, 2) in population $j=2$. This scenario represents the configuration of two populations, each population composed of 20 subjects, where each subject is composed of 100 items that are clustered in five groups. Notice that in this scenario, the item ${\bf x}_{i,2,1}$, which belongs to cluster with center at $(0, 0)$ in population $j=1$, belongs to cluster with center at $(2,2)$ in population $j=2$. Therefore, clusters with centers at (0, 0) and (2, 2) in population $j=2$ have 19 and 21 items, respectively (clusters with centers at (4, 4), (6, 6), and (8, 8) have 20 items each one), while the subjects of population $j=1$ have each cluster composed of 20 items (Figure \ref{fig:simulacao}b).

\item Scenario (iii): items of the $i$th ($i=1,\ldots,20$) subject taken from the $j$th ($j=1,2$) population are generated in the same manner as in scenario (i), except by items ${\bf x}_{i,2,1}$ and ${\bf x}_{i,2,21}$, which are centered at positions (2, 2) and (0, 0) in population $j=2$, respectively. This scenario represents the configuration of two populations, each population composed of 20 subjects, where each subject is composed of 100 items that are clustered in five groups. In this scenario, the item ${\bf x}_{i,2,1}$, which belongs to cluster with center at $(0, 0)$ in population $j=1$, belongs to cluster with center at $(2,2)$ in population $j=2$, and item ${\bf x}_{i,2,21}$, which belongs to cluster with center at $(2, 2)$ in population $j=1$, belongs to cluster with center at $(0, 0)$ in population $j=2$. Notice that, differently from scenario (ii), there is no change in the number of items in each cluster between populations, i.e., each cluster is composed of 20 items (Figure \ref{fig:simulacao}c).

\item Scenario (iv): items of the $i$th ($i=1,\ldots,20$) subject taken from the $j$th ($j=1,2$) population are generated in the same manner as in scenario (i), except by items (${\bf x}_{i,2,81}\ldots {\bf x}_{i,2,100}$) that are centered at position (6, 6). This scenario represents the change in the number of clusters between populations. Notice that population $j=1$ is composed of five clusters while population $j=2$ is composed of four clusters (items (${\bf x}_{i,2,81}\ldots {\bf x}_{i,2,100}$) belong to the same cluster of items (${\bf x}_{i,2,61}\ldots {\bf x}_{i,2,80}$)). 

\end{enumerate}

%%%%%%%%%%%%%%%%%%%%
%% Figure
%%%%%%%%%%%%%%%%%%%%

Moreover, in the following we also assume that experimental data are usually mixtures of subjects of different populations, i.e., one population is contaminated with subjects of another population and vice-versa. In order to verify the power of ANOCVA under this condition, subjects are mixed at different proportions, from 0\% (no mixture) to 50\% (half of the subjects are from one population and another half from another population, i.e., totally mixed data sets).

\begin{figure}[!t]
\centering
\includegraphics[angle=0, width=5.5in]{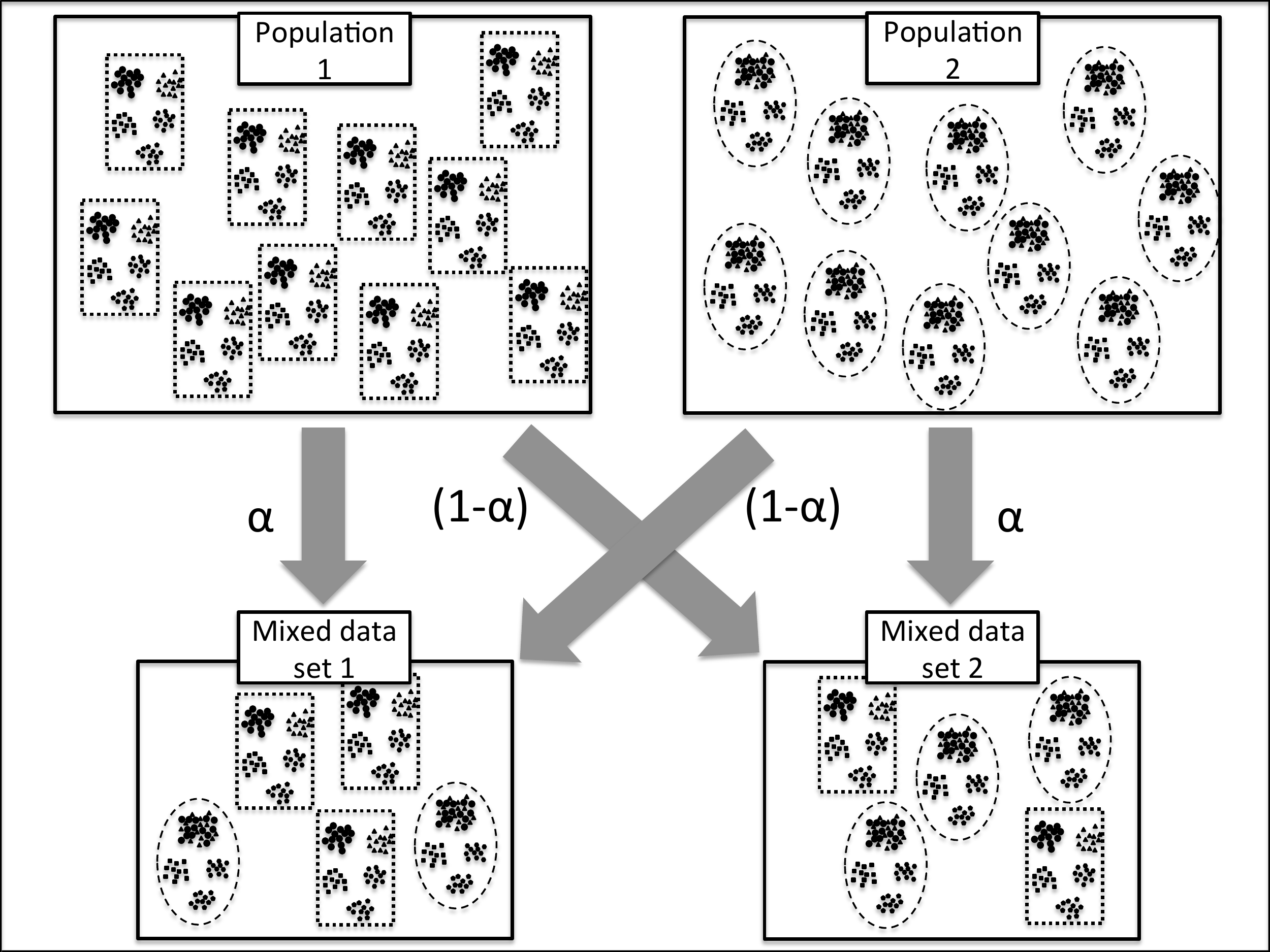}
\caption{{\bf Schema of the construction of mixed data sets.} One hundred subjects of each population ($j=1, 2$) are generated. Then, $\alpha \times 20$ subjects are randomly sampled from the 100 subjects of population $j=1$ and $(1-\alpha) \times 20$ subjects are randomly sampled from the 100 subjects of population $j=2$. The second mixed data set is constructed by sampling $\alpha \times 20$ subjects from the 100 subjects of population $j=2$ and $(1-\alpha) \times 20$ subjects from the 100 subjects of population $j=1$. The statistical test is applied to these mixed data sets 1 versus 2.}
\label{fig:simulacao-mix}
\end{figure}

In order to construct a mixed data set, 100 subjects are generated for each population ($j=1, 2$). Then, $\alpha \times 20$ subjects are randomly (uniformly) sampled from the 100 subjects of population $j=1$ and $(1-\alpha) \times 20$ subjects are randomly sampled from the 100 subjects of population $j=2$. The second data set is constructed by sampling $\alpha \times 20$ subjects from the 100 subjects of population $j=2$ and $(1-\alpha) \times 20$ subjects randomly sampled from the 100 subjects of population $j=1$ (Figure \ref{fig:simulacao-mix}). The parameter $\alpha$ varies from 0 to 0.5, where 0 means no mixture of populations and 0.5 means totally mixed data sets. ANOCVA is applied to these two mixed data sets. The clustering algorithm and the dissimilarity measure used to aforementioned simulations are the complete linkage hierarchical clustering procedure and the Euclidian distance, respectively.

%%%%%%%%%%%%%%%%%%%%
%% Figure
%%%%%%%%%%%%%%%%%%%%

\subsection{ADHD data description}\label{data-description}
ANOCVA was applied to a functional magnetic resonance imaging (fMRI) data set composed of children with typical development (TD) and with ADHD, both under a resting state protocol, totaling 759 subjects. This data set is available at the ADHD-200 Consortium \citep{ADHD200} website (\href{http://fcon\_1000.projects.nitrc.org/indi/adhd200/}{http://fcon\_1000.projects.nitrc.org/indi/adhd200/}). fMRI data was collected in eight sites that compose the ADHD-200 consortium, and was conducted with local Internal Review Board approval, and also in accordance with local Internal Review Board protocols. %All data distributed via the International Neuroimaging Data-sharing Initiative were fully anonymized in compliance with the HIPAA Privacy Rules.
Further details about this data set can be obtained at the ADHD-200 consortium website.

This data set is composed of 479 controls - children with TD (253 males, mean age $\pm$ standard deviation of 12.23 $\pm$ 3.26 years) and three sub groups of ADHD patients: (i) combined - hyperactive/impulsive and inattentive - (159 children, 130 males, 11.24 $\pm$ 3.05 years), (ii) hyperactive/impulsive (11 subjects, 9 males, 13.40 $\pm$ 4.51 years), and (iii) inattentive (110 subjects, 85 males, 12.06 $\pm$ 2.55 years).

\subsection{ADHD data pre-processing}\label{pre-processing}
The pre-processing of the fMRI data was performed by applying the Athena pipeline \\(\href{http://www.nitrc.org/plugins/mwiki/index.php/neurobureau:AthenaPipeline}{http://www.nitrc.org/plugins/mwiki/index.php/neurobureau:AthenaPipeline}). The pre-processed data is publicly available at the Neurobureau website \\(\href{http://neurobureau.projects.nitrc.org/ADHD200}{http://neurobureau.projects.nitrc.org/ADHD200}). Briefly, the steps of the pipeline are as follows: exclusion of the first four scans; slice timing correction; deoblique dataset; correction for head motion; masking the volumes to discard non-brain voxels; co-registration of mean image to the respective anatomic image of the children; spatial normalization to MNI space (resampling to 4 mm $\times$ 4 mm $\times$ 4 mm resolution); removing effects of WM, CSF, head motion (6 parameters) and linear trend using linear multiple regression; temporal band-pass filter (0.009 $<$ f $<$ 0.08 Hz); spatial smoothing using a Gaussian filter (FWHM = 6mm). The CC400 atlas (based on the approach described in Craddock {\it et al}., (2012)\citep{Craddock12}) provided by the Neurobureau was used to define the 351 regions-of-interest (ROIs) used in this study. The average signal of each ROI was calculated and used as representative of the region.

For each child, a correlation matrix was constructed by calculating the Spearman's correlation coefficient (which is robust against outliers and suitable to identify monotonic non-linear relationships) among the 351 ROIs (items) in order to identify monotonically dependent ROIs. Then, the correlations' matrices were corrected for site-effects by using a general linear model. Site effects were modeled as a GLM (site as a categorical variable) and the effect was removed by considering the residuals of this model. P-values corresponding to the Spearman's correlation for each pair of ROIs were calculated. Then, the obtained p-values were corrected by False Discovery Rate (FDR) \citep{Benjamini95}. Thus, the dissimilarity matrices ${\bf A}_{j}$ for $j=1,\ldots, 759$ are symmetric with diagonal elements equal zero and non-diagonal elements ranging from zero to one. The higher the correlation, the lower the p-value, and consequently, the lower the dissimilarity between two ROIs. Notice that the p-value associated to each Spearman's correlation is not used as a statistical test, but only as a measure of dissimilarity normalized by the variance of the ROIs. The choice of the proposed dissimilarity measure instead of the standard one minus the correlation coefficient is due to the fact that we are interested in ROIs that are highly correlated, independent whether they present positive or negative correlation. Here, we are interested in calculating how much ROIs are dependent (the dissimilarity among them) and not how they are correlated. %Moreover, positive and negative signs only make sense in monotonically dependent data. In non-monotonic relationships, the sign has no interpretation. 

\section{Results}
\label{results}

\subsection{Simulation analysis}\label{simulation-results}
\begin{figure}[!t]
\centering
\includegraphics[angle=0, width=5.5in]{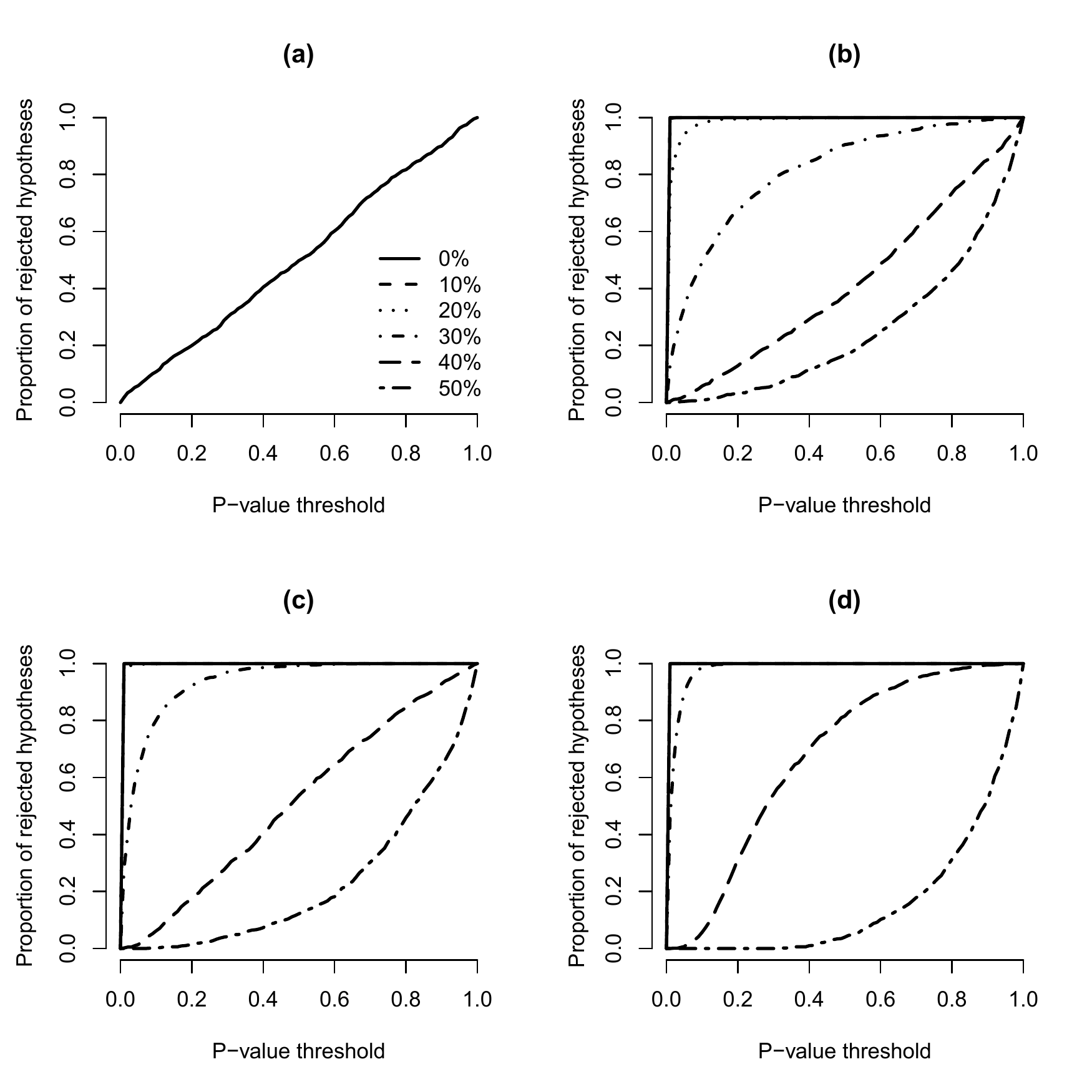}
\caption{{\bf Power curve.} The x-axis represents the p-value threshold and the y-axis represents the proportion of rejected null hypotheses in {1\,000} repetitions. Panels (a), (b), (c), and (d) represent the power curves obtained in the simulations of scenarios (i), (ii), (iii), and (iv), respectively. Notice that for low rates ($\alpha\le$30\%) of mixture, the number of truly rejected null hypotheses is high, while for higher rates of mixture ($\alpha\ge$40\%), the statistical test does not reject the null hypothesis.}
\label{fig:roc}
\end{figure}

For each scenario and each value of $\alpha$ ($\alpha=0, 0.1, 0.2, 0.3, 0.4, 0.5$), {$1\,000$} Monte Carlo realizations were constructed and tested by our approach. The results obtained by simulations are illustrated in Figure \ref{fig:roc}, which describes the proportion of rejected null hypotheses for each p-value threshold (significance level). Under the null hypothesis (scenario (i)), Figure \ref{fig:roc}a shows that the test is actually controlling the rate of type I error, i.e., the proportion of falsely rejected null hypotheses is the expected by the p-value threshold. Since the uniform distribution for p-values implies that the distribution of the statistic is correctly specified under the null hypothesis, the Kolmogorov-Smirnov test was applied to compare the p-values' distribution with a uniform distribution. The test under the null hypothesis presented a p-value of 0.14, meaning that there is no statistical evidence to affirm that the Monte Carlo p-value distribution is not a uniform distribution.
 
In the case that there is no mixture of populations ($\alpha=0$) and under the alternative hypothesis, i.e., scenarios (ii), (iii), and (iv), the test identified 100\% of the times (at a p-value threshold of 0.05), the differences in the clustering structures (Figure \ref{fig:roc}, panels (b), (c), and (d)). As the coefficient $\alpha$ of mixture level increases, the power of the test decreases, and, as expected, when the mixture level is 50\% ($\alpha=0.5$), the method is not able to identify any difference between populations (at a p-value threshold of 0.05). Moreover, by comparing the different scenarios under the same $\alpha$, it is possible to verify that the power of the test is higher in scenario (iv), where the number of clusters changes (Figure \ref{fig:roc}d), followed by scenario (iii), which represents a ``swap'' of items between clusters (Figure \ref{fig:roc}c) and finally by the case (scenario (ii)) that one item ``jumps'' from one cluster to another (Figure \ref{fig:roc}b). These results are in accordance to the intuitive notion that the power of the test is proportional to the number of items that are clustered in a different manner between populations. The greater the number of items clustered in a different manner, the higher the power of the test to discriminate it.

%%%%%%%%%%%%%%%%%%%%
%% Figure
%%%%%%%%%%%%%%%%%%%%

\begin{figure}[!t]
\centering
\includegraphics[angle=0, width=5.5in]{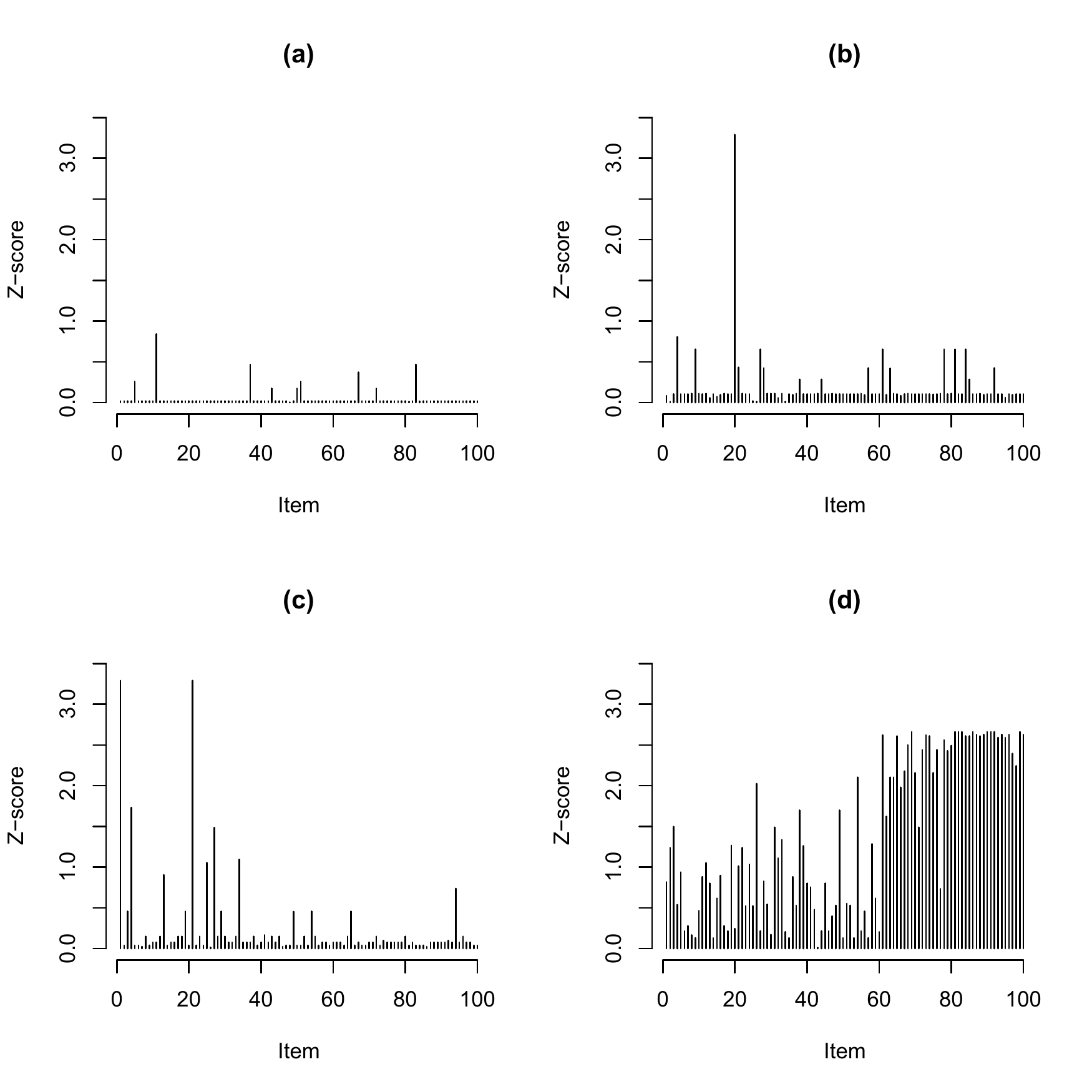}
\caption{{\bf Statistically significant items.} Panels (a), (b), (c), and (d) represent how much the items 1 to 100 contribute to the differential clustering in scenarios (i), (ii), (iii), and (iv), respectively. Items with z-score higher than 1.96 represent the statistically significant items at a p-value threshold of 0.05 after FDR correction for multiple comparisons. Notice that, as expected, panel (a) does not present any items as statistically significant. Panel (b) presents the 20th item as statistically significant. Panel (c) shows that items 1 and 21 are statistically significant. Panel (d) shows that there is a concentration of high z-scores between items 60 and 100 as statistically significant. Items that presented the highest z-scores are the ones that clustered in a different manner among populations.}
\label{fig:z-score}
\end{figure}

Figure \ref{fig:z-score} depicts an illustrative example of one realization of each scenario with $\alpha=0$ in order to show that the method indeed identifies the items that are contributing to the differential clustering among different populations. The x-axis represents the items from 1 to 100. The y-axis represents the z-scores of the p-values corrected for multiple comparisons by the False Discovery Rate (FDR) method \citep{Benjamini95}. Figure \ref{fig:z-score}, panels (a), (b), (c), and (d), represent the items and the respective z-scores that contribute significantly to the differential clustering in scenarios (i), (ii), (iii), and (iv), respectively. Items with z-score higher than 1.96 represent the statistically significant ones at a p-value threshold of 0.05. Notice that, as expected, Figure \ref{fig:z-score}a does not present any items as statistically significant because scenario (i) was constructed under the null hypothesis. Figure \ref{fig:z-score}b highlights the 20th item as statistically significant, which is exactly the item that ``jumped'' from one cluster to another. Figure \ref{fig:z-score}c shows that items 1 and 21 are statistically significant. Items 1 and 21 are the ones that were ``switched'' in our simulations. Figure \ref{fig:z-score}d illustrates a concentration of high z-scores between items 60 and 100, representing the items that were merged into one cluster in population $j=2$ of scenario (iv).

%%%%%%%%%%%%%%%%%%%%
%% Figure
%%%%%%%%%%%%%%%%%%%%

Therefore, by analyzing the statistic $\Delta s_{q}$ for $q=1,\ldots, N$, it is also possible to identify which item is contributing to the differential clustering among populations, i.e., which item is clustered in a different manner among populations.

Another point to be analyzed is that, in practice, the number of clusters should be estimated. The problem is that there is no consensus in the literature regarding this problem. In other words, different methods may estimate different number of clusters for the same data set due to the lack of definition of a cluster (a cluster is usually a result obtained by a clustering algorithm).

\begin{figure}[!t]
\centering
\includegraphics[angle=0, width=5.5in]{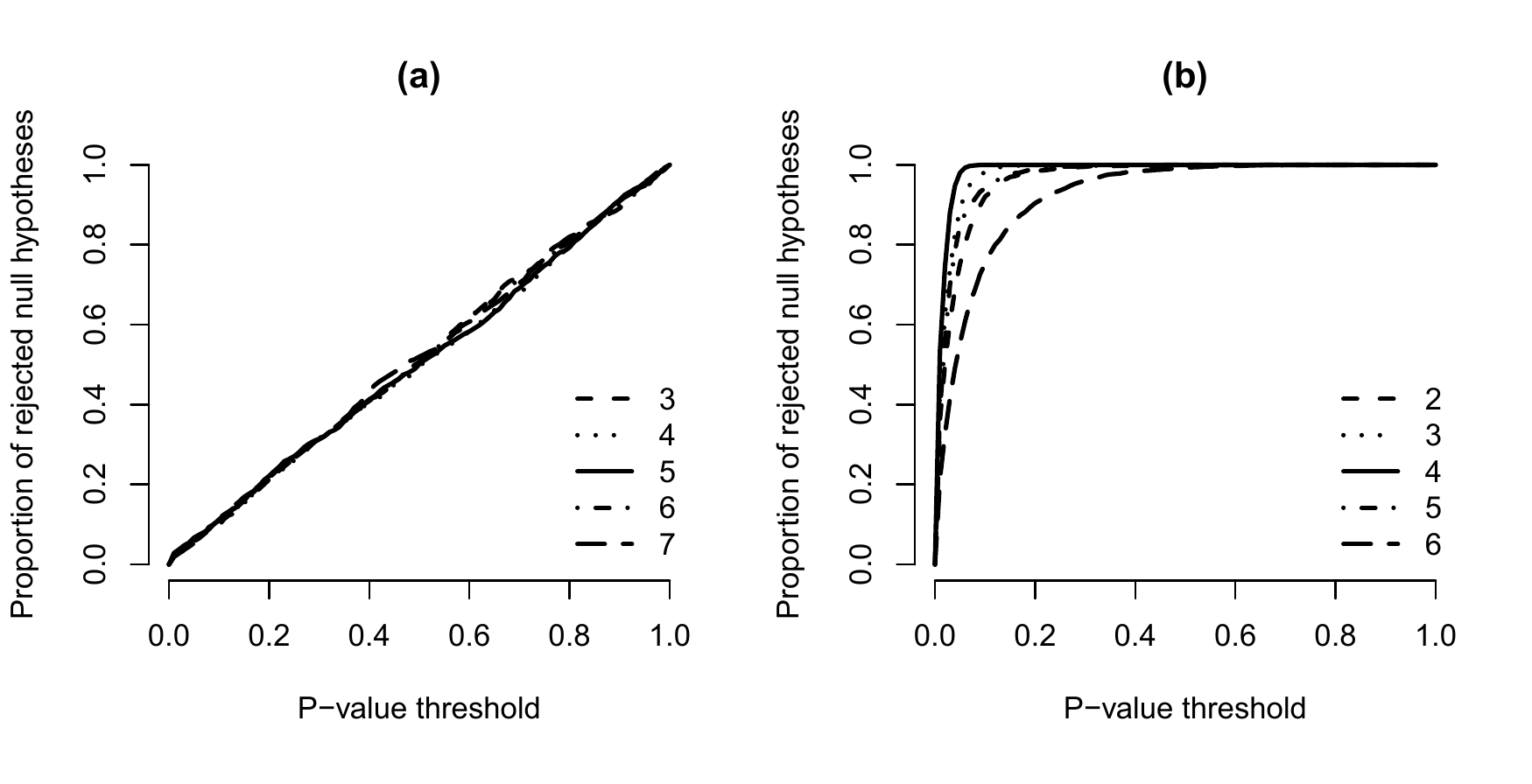}
\caption{{\bf Power curves for different number of clusters.} The x-axis represents the p-value threshold and the y-axis represents the proportion of rejected null hypotheses in {1\,000} repetitions. (a) Simulation under the null hypothesis. The correct number of clusters is five. The proportion of rejected null hypothesis is as expected by the p-value threshold. (b) Simulation under the alternative hypothesis. The correct number of clusters is four. Notice that the highest power of the test is obtained when the correct number of clusters is used.}
\label{fig:k-estimation}
\end{figure}

Then, we also analyzed how sensitive is the power of ANOCVA regarding the estimated number of clusters. In order to illustrate it, two further simulations were carried out. First, the control of the rate of false positives in scenario (i) was verified by varying the number of clusters from three to seven. The number of repetitions was set to {1\,000}. Notice that the correct number of clusters of $\bar{\bar{{\bf A}}}$ is five. Figure \ref{fig:k-estimation}a shows that the rate of rejected null hypothesis is proportional to the expected by the p-value threshold for number of clusters varying from three to seven. This result suggests that for numbers of clusters close to the ``correct'' one, the test is able to control the type I error under the null hypothesis. Now, it is necessary to verify the power of the test under the alternative hypothesis with different number of clusters. The simulation to evaluate it is composed of two populations ($j=1,2$). Each population is composed of $n_{j}=20$ subjects, and each subject is composed of $N=80$ items. These $N=80$ items are generated by normal distributions with unit variance in the following manner. For population $j=1$, items (${\bf x}_{i,1,1},\ldots, {\bf x}_{i,1,20}$), (${\bf x}_{i,1,21},\ldots, {\bf x}_{i,1,40}$), (${\bf x}_{i,1,41},\ldots, {\bf x}_{i,1,60}$), and (${\bf x}_{i,1,61},\ldots, {\bf x}_{i,1,80}$) (for $i=1, \ldots, 20$) are centered at positions (2, 0), (0, -2), (-2, 0), and (0, 2), respectively. For population $j=2$, items (${\bf x}_{i,2,1},\ldots, {\bf x}_{i,2,20}$), (${\bf x}_{i,2,21},\ldots, {\bf x}_{i,2,40}$), (${\bf x}_{i,2,41},\ldots, {\bf x}_{i,2,60}$), and (${\bf x}_{i,2,61},\ldots, {\bf x}_{i,2,80}$) (for $i=1, \ldots, 20$) are centered at positions (4, 0), (0, -4), (-4, 0), and (0, 4), respectively. The number of repetitions and the mixture tuning variable $\alpha$ are set to {1\,000} and 0.3, respectively. This simulation describes the scenario that the number of clusters for the matrix of dissimilarities $\bar{\bar{{\bf A}}}$ is clearly four. Figure \ref{fig:k-estimation}b shows that the power of the test is optimum when the correct number of clusters is used.

%%%%%%%%%%%%%%%%%%%%
%% Figure
%%%%%%%%%%%%%%%%%%%%

Thus, these results suggest that the method is still able to reject the null hypothesis with a considerable power under the alternative hypothesis for number of clusters close to the ``correct'' number. Therefore, the choice of an objective criterion to determine the number of clusters does not change significantly the results. We do not enter in further discussions regarding the estimation of the number of clusters because it is not the scope of this work. For a good review regarding the estimation of the number of clusters, refer to Milligan and Cooper (1985) \citep{Milligan85}.

\begin{figure}[!t]
\centering
\includegraphics[angle=0, width=5.5in]{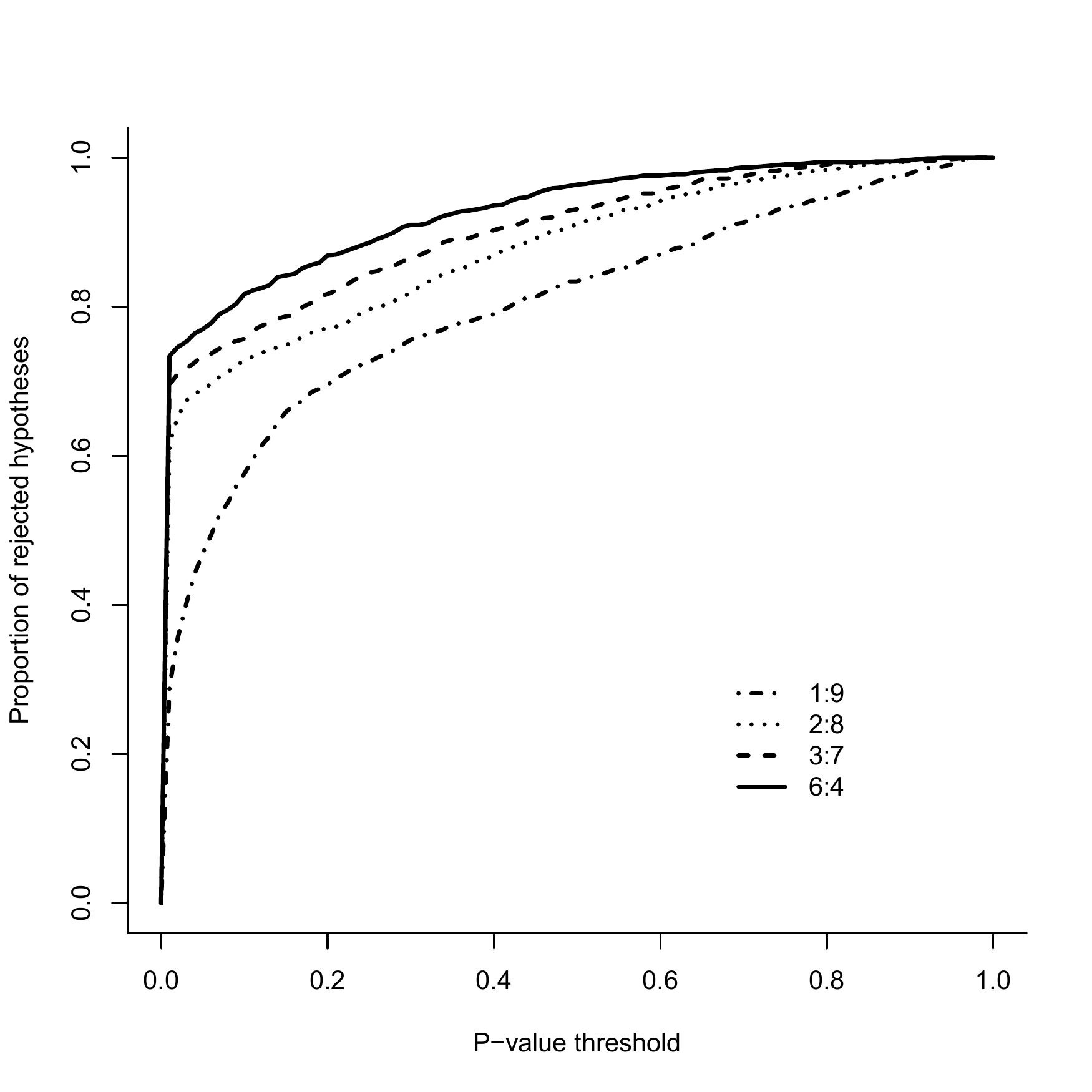}
\caption{{\bf Power curves for different balance proportions.} The x-axis represents the p-value threshold and the y-axis represents the proportion of rejected null hypotheses in {1\,000} repetitions. Notice that the more balanced are the populations, the higher is the power of the test.}
\label{fig:balance}
\end{figure}

Sometimes, populations are not balanced in their respective sizes and consequently, the largest population may dominate the average, and the clustering algorithm may bias the assignment. In order to study the performance of ANOCVA in not well balanced populations, we performed the simulation described in scenario 2 with populations varying in proportions of 1:9, 2:8, 3:7, and 4:6 in a total of 40 subjects. Power curves are shown in Figure \ref{fig:balance}. One thousand repetitions were done for each analyzed proportion. By analyzing Figure \ref{fig:balance}, one may notice that the power of the test is high when populations are closer to 5:5 and is low when they are not balanced. In other words, the more balanced are the populations, the higher is the power of the test. Besides, even when data is poorly balanced (1:9), the area under the curve is greater than 0.5.

%%%%%%%%%%%%%%%%%%%%
%% Figure
%%%%%%%%%%%%%%%%%%%%

Another point is the definition of what a cluster is. Since the definition of a cluster depends on the problem of interest - clusters are usually determined as the result of the application of a clustering algorithm (for example, k-means, hierarchical clustering, spectral clustering, etc) - it is natural that the results obtained by ANOCVA may change in function of the clustering procedure and the chosen metric (for example, euclidean, manhattan, etc). Thus, the selection of both clustering algorithm and metric depends essentially on the type of data, what is clustered, and the hypothesis to be tested.

\subsection{ADHD data analysis}\label{adhd-results}
ANOCVA was applied to the ADHD data set, in order to identify ROIs associated with the disease. Since we are interested in identifying ROIs that are differentially clustered in terms of their connectivity, the clustering algorithm used to determine the labels based on the dissimilarity matrix $\bar{\bar{{\bf A}}}$ was the unnormalized spectral clustering algorithm \citep{Ng02}. For details of the implemented spectral clustering algorithm, refer to the Appendix.  
The number of clusters was determined by using the silhouette method \citep{Rousseeuw87}. The number of bootstrap samples was set to {1\,000}. The group of children with hyperactive/impulsive ADHD was excluded from our analysis due to the low number of subjects (11 children). The performed tests and the respective p-values corrected for multiple comparisons by the Bonferroni method are listed at Table \ref{table:adhd}. First, the test was applied to the entire data set (excluded the group of hyperactive/impulsive ADHD due to the low number of subjects) in order to verify if there is at least one population that differs from the others. The test indicated a significant difference (p-value=0.020), suggesting that there is at least one population that presents a different clustering structure. In order to identify which populations present different clustering structures, pairwise comparisons among the groups were carried out. By observing Table \ref{table:adhd}, it is not possible to verify significant differences between TD versus inattentive ADHD (p-value = 0.700), and combined ADHD versus inattentive ADHD (p-value = 0.615), but there are significant differences between TD versus combined and inattentive ADHD (p-value $<$ 0.001), and TD versus combined ADHD (p-value $<$ 0.001). These results indicate that the significant differences obtained for TD, combined ADHD and inattentive ADHD were probably due to the differences between TD and combined ADHD.

%%%%%%%%%%%%%%%%%%%%
%% Table
%%%%%%%%%%%%%%%%%%%%
\begin{table}%[ht]
\caption{ANOCVA applied to the ADHD data set. The number of bootstrap samples is set to {1\,000}. P-values are corrected by Bonferroni method for multiple comparisons.}
\centering
\begin{tabular}{c c}
\hline\hline
Comparison & P-value\\
\hline
TD vs Combined ADHD vs Inattentive ADHD & 0.020\\
TD vs Combined ADHD and Inattentive ADHD & $<$0.001\\
TD vs Combined ADHD & $<$0.001\\
TD vs Inattentive ADHD & 0.700\\
Combined ADHD vs Inattentive ADHD & 0.615\\
\hline
\hline
\end{tabular}
\label{table:adhd}
\end{table}

\begin{figure}[!t]
\centering
\includegraphics[angle=0, width=5.5in]{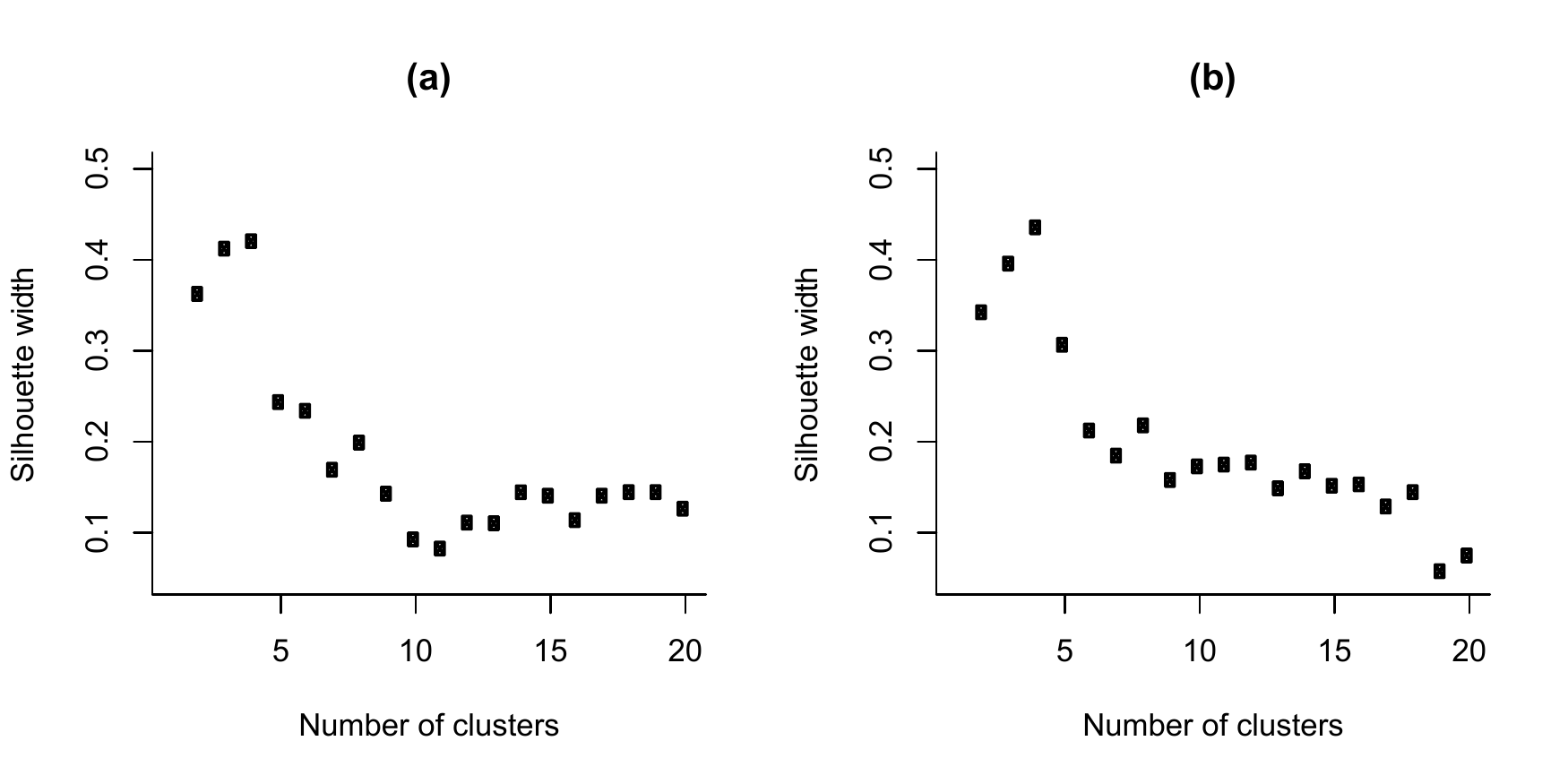}
\caption{{\bf Selection of the number of clusters.} The number of clusters versus the silhouette width (the silhouette widths for children with TD and with combined ADHD are defined as $\frac{1}{479}\sum_{q=1}^{479}s_{q}^{(\bar{\bf{A}}_{\text{\bf{TD}}},{\bf l}_{\bar{\bf {A}}_{\text{TD}}})}$ and $\frac{1}{159}\sum_{q=1}^{159}s_{q}^{(\bar{\bf{A}}_{\text{\bf{combined ADHD}}},{\bf l}_{\bar{\bf {A}}_{\text{combined ADHD}}})}$, respectively). (a) Estimation of the number of clusters in the average dissimilarity matrix $\bar{{\bf A}}_{\text{TD}}$ of an fMRI data set composed of 479 children with TD. (b) Estimation of the number of clusters in the average dissimilarity matrix $\bar{{\bf A}}_{\text{combined ADHD}}$ of an fMRI data set composed of 159 children with combined ADHD. Notice that the silhouette width for one cluster is not defined. The maximum silhouette width is obtained when the number of clusters is four for both cases. These results suggest that the number of clusters in both data sets is four.}
\label{fig:silhouette}
\end{figure}

Thus, the analysis of the fMRI data set was focused on identifying the differences between children with TD and children with combined ADHD. ROIs of both populations were clustered by spectral clustering algorithm, and the numbers of clusters for each group were estimated by the silhouette method. The silhouette method estimates the optimum number of clusters by selecting the number of clusters associated with the highest silhouette width. By analyzing Figure \ref{fig:silhouette}, it is possible to see that the estimated number of clusters is four for both data sets.

%%%%%%%%%%%%%%%%%%%%
%% Figure
%%%%%%%%%%%%%%%%%%%%

\begin{figure}[!t]
\centering
\includegraphics[angle=0, width=5.5in]{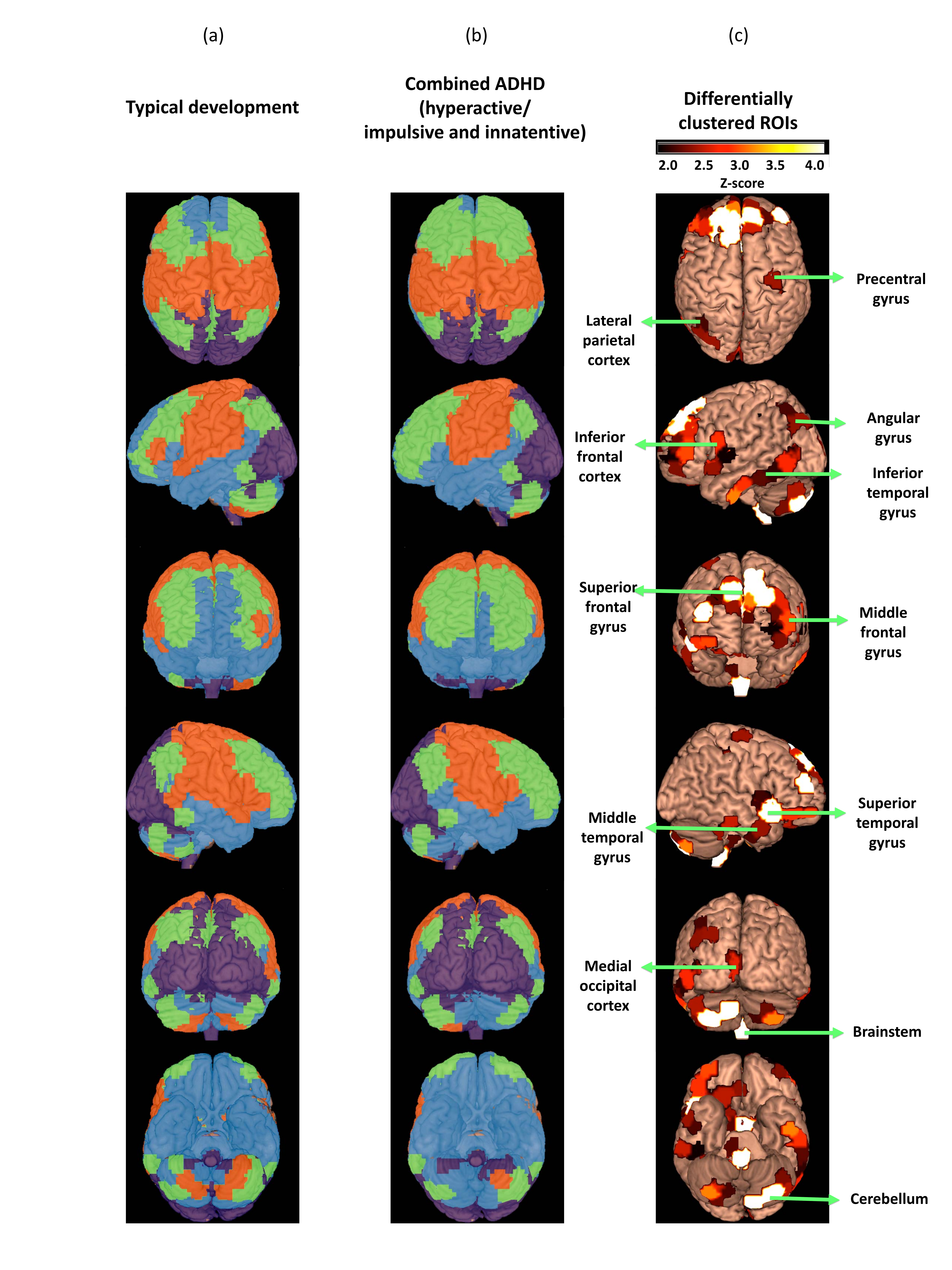}
\caption{{\bf Clustering of ROIs.} Panels (a) and (b) represent the ROIs clustered by the spectral clustering algorithm applied to the dissimilarity matrices $\bar{{\bf A}}_{\text{TD}}$ and $\bar{{\bf A}}_{\text{combined ADHD}}$, respectively. The number of clusters was estimated by the silhouette method. Panel (c) highlights the ROIs that are clustered in a different manner between children with TD and ADHD. Regions highlighted in white represent high z-scores while regions in red represent lower z-scores. The z-scores were calculated by using the p-values obtained by ANOCVA after FDR correction for multiple comparisons.}
\label{fig:brain}
\end{figure}

The results of the clustering procedure are visualized in Figure \ref{fig:brain}, where panels (a) and (b) represent children with TD and children with combined ADHD, respectively.

%%%%%%%%%%%%%%%%%%%%
%% Figure
%%%%%%%%%%%%%%%%%%%%

Interestingly, the clusters were composed of anatomically contiguous and almost symmetric areas of the brain, although these constraints were not included {\it a priori} in our analyses. This is consistent with the hypothesis that the spectral clustering method groups areas with similar brain activities in the same cluster.

Then, each ROI was tested in order to identify the one that significantly contribute to the difference in clustering between children with TD and with combined ADHD. P-values were corrected for multiple comparisons by the FDR method, and then, converted to z-scores. Figure \ref{fig:brain}c illustrates the statistically significant ROIs at a p-value threshold of 0.05 after FDR correction. The regions highlighted in white are the ROIs with the highest z-scores, while the regions highlighted in red represent ROIs with lower z-scores, but still statistically significant.

By comparing Figure \ref{fig:brain} panels (a) and (b), it is possible to verify that the highlighted regions in Figure \ref{fig:brain}c correspond to ROIs that are clustered in a different manner between children with TD and with combined ADHD.

Cluster analysis has suggested a very similar network organization between children with TD and combined ADHD patients. Apparently, sensory-motor systems, frontoparietal control networks, visual processing and fronto-temporal systems are similarly distributed between the two groups. However, the application of ANOCVA unveiled that anterior portion of inferior, middle and superior frontal gyri, inferior temporal gyrus, angular gyrus, and some regions of cerebellum, lateral parietal, medial occipital and somato-motor cortices have a distinct clustering organization between the two populations.

Motor system (pre and postcentral giri and cerebellum) alterations in ADHD are associated to hyperactivity symptoms, and this finding has already been extensively described in the literature (for a detailed review, see \citep{Castellanos12}). In addition, Dickstein {\it et al}.\citep{Dickstein06} carried out a meta-analysis of fMRI studies comparing controls and ADHD and identified some portions of parietal cortex, inferior prefrontal cortex and primary motor cortex as regions with activation differences between the groups.

The inferior frontal cortex highlighted by ANOCVA is described in literature as a key region for inhibition of responses \citep{Aron04}.  In this sense, the impulsivity symptoms present in combined ADHD can be related to an abnormal participation of this region in the context of global brain network organization, when compared to healthy controls. This finding is reinforced by the findings of recent studies. Schulz {\it et al}. \citep{Schulz12} investigated the role of this area in therapeutic mechanisms of treatments for ADHD. In addition, Vasic {\it et al}. \citep{Vasic12} and Whelan {\it et al}. \citep{Whelan12} explored the neural error signaling in these regions (in adults) and the impulsivity of adolescents with ADHD. The inferior frontal cortex is also implied to participate in language production, comprehension, and learning and, therefore, our finding is consistent with the language impairment reported in ADHD subjects \citep{Tirosh98}.

An interesting finding from the application of the proposed method to the resting state fMRI dataset was the identification of angular gyrus as a region with functional abnormalities in ADHD in the context of brain networks. Although the angular gyrus contributes to the integration of information, playing an important role in many cognitive processes \citep{Seghier12}, to the best of our knowledge, there are very few studies in literature suggesting activation differences in this region when comparing ADHD to healthy controls \citep{Seghier12, Tamm06}. The angular gyrus contributes to the integration of information, and play a role in many cognitive processes \citep{Seghier12}. Tamm {\it et al}. \citep{Tamm06} have found that this region exhibited less activation in adolescents with ADHD during a target detection task. Moreover, Simos {\it et al}. \citep{Simos11} have shown that angular and supramargial gyrus play a role in brain mechanisms for reading and its correlates in ADHD. We note that despite the angular gyrus has a crucial role in the integration of information, both studies have not explored the relevance of this region from a network connectivity perspective. Using the proposed method, we properly carried out this analysis and the findings indicate that the role of this region in brain spontaneous activity is different, when comparing the two groups.

%Finally, it is important to mention that the proposed method was able to identify several differences in distributed regions across the brain in a single analysis, when most studies in literature often present more “focal” differences between ADHD patients and typical developing subjects.

The existence of temporal and spatial correlation is inherent in fMRI data and ignoring intrinsic correlation may lead to misleading or erroneous conclusions. This dependence structure makes clustering analysis more challenging and should be accounted for \citep{Kang12, Liao08}. Notice that the proposed bootstrap incorporates the spatial correlations in the clustering process and also preserves the temporal structure.

In order to certify that both the bootstrap based statistical test is corretly working in actual data and the results obtained in this study are not due to numerical fluctuations or another source of error not took into account, we verified the control of the rate of false positives in biological data. The set of 479 children with TD was split randomly into two subsets, and the clustering test was applied between them. This procedure was repeated 700 times. The proportion of falsely rejected null hypothesis for p-values lower than 1, 5, and 10\% were 2.14\%, 5.70\%, and 9.83\%, respectively, confirming that the type I error is effectively controlled in this biological data set. Moreover, the Kolmogorov-Smirnov test was applied to compare the p-values' distribution obtained in the 700 repetitions with the uniform distribution. The Kolmogorov-Smirnov test indicated a p-value of 0.664, meaning that there is no statistical evidence to reject the null hypothesis that the p-values' distribution obtained in 700 repetitions follows a uniform distribution.

Furthermore, we also verified in the same manner whether the highlighted ROIs are indeed statistically significant. The proposed method was applied to each ROI, i.e., 351 p-values were calculated in each repetition. Thus, 351 p-values' distributions, one for each ROI were constructed. Each of the 351 p-values' distribution was compared with the uniform distribution by the Kolmogorov-Smirnov test. After correcting the obtained 351 p-values by the Kolmogorov-Smirnov test by FDR \citep{Benjamini95}, only two null hypotheses were rejected at a p-value threshold of 0.05, confirming that the type I error is also controlled for ROIs. 

These results suggest that the differences in clustering between children with TD and with combined ADHD are indeed statistically significant.

\section{Final remarks}\label{final}
To the best of our knowledge, the method proposed here is the first one that statistically identifies differences in the clustering structure of two or more populations of subjects simultaneously.

However, it is important to discuss some limitations of ANOCVA. First, the method is only defined if the estimated number of clusters for the average of the matrix of dissimilarities $\bar{\bar{{\bf A}}}$ is greater than one. It is due to the fact that the silhouette statistic $s$ is only defined when the number of clusters is greater than one. In practice, one may test whether each dissimilarity matrix $\bar{{\bf A}}_{j}$ and the average of dissimilarity matrix $\bar{\bar{{\bf A}}}$ are composed of one or more clusters by using the gap statistic proposed by Tibshirani {\it et al}. (2002)\citep{Tibshirani02}. If $\bar{\bar{{\bf A}}}$ is composed of one cluster while one of the matrices of dissimilarities $\bar{{\bf A}}_{j}$ is composed of more than one cluster, clearly the clustering structures among populations are different.

Another limitation is the fact that ANOCVA does not identify changes in both rotated and/or translated data. In other words, the test does not identify alterations that maintain the relative dissimilarities among items. If one is interested in identifying this kind of difference, one simple solution is to test the joint mean by the Hotteling's T-squared test \citep{Hotteling31}.

However, it is important to point out that ANOCVA is sensitive to identify different clustering assignments that have the same goodness of fit (silhouette). Notice that \cite{Rousseeuw87} proposed the use of the average of $s_q$ in order to obtain a goodness of fit. Here, we do not use the average value but the distance between the entire vectors $s_q^{(\bar{\bar{{\bf A}}},{\bf l}_{\bar{\bar{{\bf A}}}})} \quad \mbox{and} \quad s_q^{(\bar{{\bf A}}_j,{\bf l}_{\bar{\bar{{\bf A}}}})}$. In other words, we take into account the label of the items that are clustered. Therefore, if one or more items are clustered in a different manner among populations, our statistic $\Delta S$ is able to capture this difference. However, the use of the average value of $s_q^{(\bar{\bar{{\bf A}}},{\bf l}_{\bar{\bar{{\bf A}}}})} \quad \mbox{and} \quad s_q^{(\bar{{\bf A}}_j,{\bf l}_{\bar{\bar{{\bf A}}}})}$ is not.

Moreover, ANOCVA requires a considerable number of subjects in each group to be able to reject the null hypothesis (when the clustering structures are in fact different). It is very difficult to define a minimum number of subjects because it depends on the variance, but we suppose that an order of dozens (based on our simulations) is necessary.

There are other measures for similarity between clustering structures \citep{Bae06, Torres08} that might be used to develop a statistical test. However, these similarity measures cannot be extended in a straightforward manner to simultaneously test more than two populations.

The work proposed by \citep{Alexander-Bloch12} was successfully applied in neuroscience to statistically test differences in network community structures. However, again, this method cannot test simultaneously more than two populations. Notice that in our case, we are interested in comparing controls and several sub-groups of ADHD simultaneously. Therefore, this method is not applicable. It is necessary to point out that the characteristic of ANOCVA that allows to statistically test whether the structure of the clusters of several populations - not limited to pairwise comparisons - are all equals avoids the increase of committing a type I error due to multiple tests. Furthermore, ANOCVA can be used to test any clustering structure, not limited to network community structures such as the one proposed by \citep{Alexander-Bloch12}.

Another advantage is the use of a bootstrap approach to construct the empirical null hypothesis distribution from the original data. It allows the application of the test to data sets that the underlying probability distribution is unknown. Moreover, it is also known that for actual data sets, the bootstrap procedure provides better control of the rate of false positives than asymptotic tests \citep{Bullmore99}.

One question that remains is, what is the difference between ANOCVA and a test for equality of dissimilarities, for example, a t-test for each distance. The main difference is that, for a t-test, only the mean and variance of the measure to be tested is taken into account to determine whether two items are ``far'' or ``close'', while ANOCVA is a data-based metric that uses the clustering structure to determine how ``far'' the items are from others. Furthermore, we also remark that testing whether the data are equally distributed is not the same of testing whether the data are equally clustered, since items may come from very different distributions and be clustered in a quite similar way depending on the clustering algorithm. One may also ask the difference between an F-test of the clustering coefficient and ANOCVA. Notice that the clustering coefficient is a measure of degree to which nodes in a graph tend to cluster together, while ANOCVA tests whether the structure of the clusters of several populations are all equals.

Here, the purpose of the test was to identify ROIs that are associated with ADHD, but the same analysis can be extended to other large data sets. Specifically in neuroscience, the recent generation of huge amounts of data in collaborative projects, such as the Autism Brain Image Data Exchange (ABIDE) project \\(\href{http://fcon\_1000.projects.nitrc.org/indi/abide/index.html}{http://fcon\_1000.projects.nitrc.org/indi/abide/index.html}), which generated fMRI data of more than {1\,000} individuals with autism, the ADHD-200 project \citep{ADHD200} previously described here that provides fMRI data of $\approx 700$ children with ADHD, the fMRI Data Center, which is a public repository for fMRI\\ (\href{http://www.fmridc.org/f/fmridc}{http://www.fmridc.org/f/fmridc}) \citep{Van2001}, the Alzheimer's Disease Neuroimaging Initiative, which collected magnetic resonance imaging of $\approx 800$ subjects \citep{Jack08},  and many others that will certainly be produced due to the decreasing costs in data acquisition, makes cluster analysis techniques indispensable to mine information useful to diagnosis, prognosis and therapy.

The flexibility of our approach that allows the application of the test on several populations simultaneously (instead of limiting to pairwise comparisons), along with its performance demonstrated in both simulations and actual biological data, will make it applicable to many areas where clustering is a source of concern.

\section{Appendix}\label{appendix}
Spectral clustering refers to a class of techniques which rely on the eigen-structure of a similarity matrix to partition points into clusters with points in the same cluster having high similarity and points in different clusters having low similarity.

The similarity matrix is provided as an input and consists of a quantitative assessment of the relative similarity of each pair of attributes in the data set. In our case, the similarity of two ROIs is given by one minus the p-value obtained by the Spearman's correlation between ROIs.

The spectral clustering algorithm is described as follows \citep{Ng02}:

{\bf Input}: The similarity matrix ${\bf H}$, where ${\bf H}=(h_{l,q})_{l,q=1,\ldots, N}$ (consider $h_{l,q}=1-\text{p-value}$ of the Spearman's correlation), and the number of clusters $r$.
\begin{enumerate}
	\item Compute the Laplacian matrix ${\bf Q}={\bf D}-{\bf H}$, where ${\bf D}$ is the diagonal matrix with $d_{1}, \ldots, d_{N}$ on the diagonal ($d_{l}=\sum_{q=1}^{N}h_{l,q}$).
	\item Compute the first $r$ eigenvectors ${\bf v}_{1}, \ldots, {\bf v}_{r}$ of ${\bf Q}$.
	\item Let ${\bf V}\in \mathbb{R}^{N\times r}$ be the matrix containing the vectors ${\bf v}_{1}, \ldots, {\bf v}_{r}$ as columns.
	\item For $l=1, \ldots, N$, let ${\bf y}_{l}\in \mathbb{R}^{r}$ be the vector corresponding to the $l$th row of ${\bf V}$.
	\item Cluster the points $({\bf y}_{l})_{l=1,\ldots,N}$ with the $k$-means algorithm into clusters $C_{1},\ldots, C_{r}$.
\end{enumerate}
{\bf Output}: Clusters $C_{1},\ldots,C_{r}$.

\section*{Acknowledgements}

AF was partially supported by FAPESP (11/07762-8) and CNPq (306319/2010-1). DYT was partially supported by Pew Latin American Fellowship. AGP was partially supported by FAPESP (10/50014-0).

\end{document}